\documentclass{article}
\usepackage{cite}
\usepackage{hhline}
\usepackage[dvips]{graphicx}

\newcommand{\bsig}{\mbox{\boldmath{$\sigma$}}}
\newcommand{\btau}{\mbox{\boldmath{$\tau$}}}
\newcommand{\br}{\mbox{${\bf r}$}}
\newcommand{\bp}{\mbox{${\bf p}$}}

\begin{document}
\title{Nuclear magnetization distribution and hyperfine splitting in
Bi$^{82+}$ ion}
\author{ R.A. Sen'kov and V.F. Dmitriev,\\
{\it Budker Institute of Nuclear Physics, Novosibirsk-90, 630090, Russia}}
\maketitle

\begin{abstract}
Hyperfine splitting in Bi$^{82+}$ and Pb$^{81+}$ ions was calculated using
continuum RPA approach with effective residual forces. To fix the
parameters of the theory the nuclear magnetic dipole
moments of two one-particle and two one-hole nuclei around $^{208}$Pb
were calculated using the same approach. The contribution
from velocity dependent two-body spin-orbit residual interaction
was calculated explicitly. Additionally, the octupole moment of
$^{209}$Bi and the hfs in muonic bismuth atom were calculated as well
in the same approach. All the calculated observables, except the electronic hfs
in $^{209}$Bi, are in good agreement with the data. We argue for more accurate
measurement of the octupole moment and the muonic hfs for $^{209}$Bi.
\end{abstract}

\section{Introduction}

High experimental precision attained in the laser spectroscopic
measurement of the ground-state hfs in hydrogen-like
${}^{209}$Bi${}^{82+}$ \cite{hfs} stimulates considerable
theoretical activity in this field (see, e.g.,
\cite{shab2,gust,labz,tom1,tom2,schn,perss,blund,pers} and
references therein).
\begin{equation} \label{exp}
\Delta E_{hfs}^{exper.} =\,5.0840(8)\, \mbox{eV}.
\end{equation}

The first quantum mechanical calculation of the hfs was made by
Fermi in $1930$ \cite{fermi}. He treated the hfs with a
non-relativistic formalism in studies of alkali atoms, where he
derived approximate generalizations from the hydrogenic case. In
the case of ${}^{209}$Bi${}^{82+}$ the formulae obtained by Fermi
give a value about $2.75$ eV, which is $46\%$ smaller than the
experimental value (\ref{exp}). Relativity can be taken into
account by multiplying non-relativistic value with a relativistic
correction factor $A(\alpha Z)$ \cite{breit}. In bismuth case $Z=83$
and factor $A(\alpha Z)=2.125$. The relativistic generalization of
the Fermi formulae holds for point-like nucleus and gives a hfs
for ${}^{209}$Bi${}^{82+}$ of about $5.84$ eV, which is $15\%$
larger than the experimental value (\ref{exp}).

The effect of spatial distribution of the nuclear charge was
analyzed in pioneering works by Rosenthal, Breit \cite{rosen} and
others \cite{crawf,strok} and it is sometimes called
"Breit-Rosenthal effect". The correction factor for the nuclear
charge distribution can be written as $(1-\delta)$, where $\delta$
is a small number which depends mainly on the root-mean-square
(rms) radius of the nuclear charge distribution,
$\left<r_c{}^2\right>^{1/2}$. By assuming a uniform spherical
symmetric charge distribution and using the experimental value
$\left<r_c{}^2\right>^{1/2} =\, 5.519(4)$ fm \cite{devri} for the
nuclear charge distribution, $\delta$ can be calculated to be
$0.0110$ in the case of ${}^{209}$Bi${}^{82+}$ \cite{shab}. The
relativistic hfs energy splitting in ${}^{209}$Bi${}^{82+}$ for a
uniformly charged nucleus becomes than about $5.20$ eV, which is
$2\%$ larger than the experimental value (\ref{exp}).

There are two principal corrections to the Fermi-Breit formulae
that is necessary to take into account: the magnetic moment
distribution within the nucleus
\cite{bweis,bohr,shab2,gust,labz,tom1,tom2} and radiative
corrections \cite{shab2,schn,perss,blund,pers}. Both of these
corrections are of comparable magnitude.

The correction for an extended nuclear magnetization was first
studied in an innovative work by A. Bohr and V. Weisskopf
\cite{bweis,bohr} and is also called "Bohr-Weisskopf effect".
Recently, three approaches for theoretical determination of the
Bohr-Weisskopf effect have been used. The simplest of these
approaches is based on a solution of the Schr\"{o}dinger equation
for a nucleon in a mean-field potential, the solution gives the
distribution for the unpaired nucleon in the nucleus and the
distribution is used to determine hyperfine splitting. This
approach has been used by Shabaev {\it et al.} \cite{shab2} and by
Gustavsson {\it et al.} \cite{gust}. A slightly more sophisticated
approach, giving equivalent results for the hfs in leading order,
is the "dynamical proton model" (DPM), where the odd proton of the
bismuth nucleus is treated as a Dirac particle bound in a
mean-field potential. The first order hfs in hydrogen-like Bi is
then given by a photon exchange between the electron and the
proton. DPM was introduced by Labzowsky {\it et al.} \cite{labz}.
The third and more complete approach is a many-body calculation
with the use of "dynamical correlation model" (DCM). This approach
is the only calculation of the Bohr-Weisskopf effect which
includes many-body contributions and Tomaselli {\it et al.} have
used it for studies of several systems \cite{tom1,tom2}.

In addition to the dominant electrostatic and hyperfine
interactions with nucleus, the electron also interacts with
radiation field, an interaction described by QED. The leading QED
corrections originate from  the one-loop self-energy and
vacuum-polarization effects. The one-loop QED effects for hfs have
been calculated by different groups and the results are consistent
\cite{shab2,schn,perss,blund,pers}.

In this work we consider a contribution of the magnetic moment
distribution within the nucleus to the hydrogen-like ion
hfs. Our approach is close to the one used in \cite{dmit}. This
microscopic approach is based mainly on the Migdal's theory of
finite Fermi system (FFST) \cite{mig} (see also
\cite{speth,shlom,sap}). Essentially, this approach is equivalent
to calculation of the core polarization effects with the use of an
effective interaction. The first calculation of the core
polarization goes back to fifties \cite{arima} and this early
development was summarized in \cite{speth}. Modern calculations,
however, differ significantly in the size of the single-particle
space included into sum over intermediate particle-hole states.
For spherical nuclei all single-particle space, including
continuum is included into calculation. This is one difference
between our approach and the DCM where one has to use the
restricted single-particle space. Another difference between our
work and \cite{tom1,tom2} is that RPA accounts for ground state correlations
represented graphically by backward going loops. And, finally, we
 use explicitly the correction
to nuclear electromagnetic current due to velocity dependent
interaction. This correction was absent in \cite{tom1,tom2} since
they use velocity independent forces. Previously, this correction
was discussed in \cite{dmit} in connection to the Bohr-Weisskopf
effect in bismuth muonic atom. It was shown that this correction
is not negligible. Here we extend this analysis to the
hydrogen-like ions around $^{208}$Pb nucleus.

\section{Basic Equations}

In FFST the effect of the core polarization is described by
introducing an effective single-particle operator (vertex) $M$
satisfying the equation \cite{mig}
\begin{equation}\label{main}
M(\omega)\,=\,e_{eff}\,M_0\,+\,F\,A(\omega)\,M(\omega),
\end{equation}
where $M_0$ is in our case the bare single-particle $MJ$ operator,
$A(\omega)$ the polarization operator of a particle-hole pair,
\begin{equation}
A^{(\omega)}_{\nu_1\,\nu_2;\nu_3\,\nu_4}\,=\,\int\,\frac{d\varepsilon}{2
\pi i}\,G_{\nu_1\,\nu_3}(\varepsilon+\frac{1}{2}\omega)\,
G_{\nu_2\,\nu_4}(\varepsilon-\frac{1}{2}\omega),
\end{equation}
where $G_{\nu\nu'}(\varepsilon)$ is the single-particle Green
function. $F$ is the amplitude of quasi-particle interaction.
Equation (\ref{main}) is written in the re-normalized form, i.e.
after elimination of a regular part in the Green function
$G_{\nu\nu'}(\varepsilon)$. The regular part of the $G$-function
corresponding to admixture of three- or more particle states produces
the effective charges $e_{eff}$, and re-normalizes the interaction amplitude
$ F$  in (\ref{main}). Following
\cite{mig}, we choose $e_{eff}$ as a constant independent of the
particular state $\left| \nu \right>$, but different for the spin
and orbital parts of the operator $M_0$:
\begin{eqnarray}
\label{zetas} e_{eff}\,M^s_0\,=\,\left\{ \begin{array}{ll}
                             (1-\zeta_s)\,(M^s_0)_p\,+\,\zeta_s\,(M^s_0)_n\,, & \mbox{for protons,}\\
                             (1-\zeta_s)\,(M^s_0)_n\,+\,\zeta_s\,(M^s_0)_p\,, & \mbox{for neutrons,}\\
                           \end{array} \right. \\
\label{zetal} e_{eff}\,M^l_0\,=\,\left\{ \begin{array}{ll}
                             (1-\zeta_l)\,(M^l_0)_p\,, & \mbox{for protons,}\\
                             \zeta_l\,(M^l_0)_p\,,     & \mbox{for neutrons.}\\
                           \end{array} \right.
\end{eqnarray}
The constants $\zeta$ in (\ref{zetas}) and (\ref{zetal}) were
taken as equal for both protons and neutrons since the main
deviation of the magnetic moments from single-particle values is
in the isovector part of the $MJ$ operator.

The effective interaction is $F=F_{ss}+F_{ls}$, where
\begin{equation}\label{ssint}
F_{ss}\,=\,C\,(g + g' \, \btau_1 \cdot \btau_2\,)\, \bsig_1 \cdot
\bsig_2\,\delta(\br_1 -\br_2)
\end{equation}
is the Migdal-type spin-spin zero-range interaction and
\begin{equation}\label{lsint}
F_{ls}\,=\,C\,r_0^2\,(\kappa + {\kappa'}\, \btau_1 \cdot
\btau_2\,) \cdot (\bp_1 -\bp_2) \cdot (\bsig_1 + \bsig_2) \times
\nabla_1 \delta(\br_1 - \br_2)
\end{equation}
is the spin-orbit one.

The $MJ$ vertex $\mathcal{M}_{JM}$ can be written in coordinate
representation in the form
\begin{equation}\label{vertex}
\mathcal{M}_{JM}\,=\,\sum_{i=1}^{3}\,v{}^i(r)\,T^{(i)}_{JM},
\end{equation}
where we have introduced the complete set of linear independent
tensor operators
\begin{equation}
\begin{array}{c}\label{operators}
T^{(1)}_{JM}\,=\,\bsig \cdot {\bf Y}^{J-1}_{JM}({\bf n}),\;
T^{(2)}_{JM}\,=\,\bsig \cdot {\bf Y}^{J+1}_{JM}({\bf n}),\\
T^{(3)}_{JM}\,=\,\frac{1}{2}\left( {\bf L} \cdot {\bf
Y}^{J-1}_{JM}({\bf n}) + {\bf Y}^{J-1}_{JM}({\bf n}) \cdot {\bf L}
\right),
\end{array}
\end{equation}
where ${\bf Y}^L_{JM}({\bf n})$ being the vector spherical
harmonic \cite{varsh}.

The values of the $MJ$ moments can be expressed in terms of matrix elements
of the vertex (\ref{vertex}):
\begin{eqnarray}
\mu_{J}\,=\,\left< \nu m=I \right| \mathcal{M}_{J0} (\omega=0)
\left| \nu m=I \right>\,=\,\left( \begin{array}{ccc}
                                    I & J & I \\
                                    -I& 0 & I
                                  \end{array} \right)\,
\sum_{i=1}^3 \,\left< n l I \right| v{}^i(r) \left| n l I \right>
\, t^{(i)}_{\nu\nu},
\end{eqnarray}
where $\nu=(n l I m)$ is the set of nucleon quantum numbers, and
$t^{(i)}_{\nu_2\nu_1}\,=\,\left( l_2 I_2 \| T^{(i)}_J \| l_1
I_1\right)$ is the reduced matrix element of the tensor operators
(\ref{operators}).

Reducing the angular and spin variables in equation (\ref{main})
we obtain in coordinate representation the system of integrals
equations
\begin{equation}\label{main2}
w{}^i(r)\,=\,e_{eff}\,w_0{}^i(r)\,+\,\sum_{j=1}^3 \int_0^\infty
dr' \Theta^{ij}(r,r'|\omega)\,w{}^j(r'),
\end{equation}
where we have introduce $w{}^i(r)\,=\,rv{}^i(r)$. The kernel of
the integrals equations (\ref{main2}) is the sum of two terms
\begin{equation} \label{kern}
\Theta^{ij}\,=\,\Theta^{ij}_{ss}\,+\,\Theta^{ij}_{ls}.
\end{equation}
For the spin-spin interaction (\ref{ssint})
\begin{equation}
\Theta^{ij}_{ss}\,=\,C \hat{g}\,\, r A^{ij}_J(r,r'|\omega) r',
\end{equation}
where $\hat{g}\,=\,g + {g'} \, \btau_1 \cdot \btau_2\,$ and the
sun over isospin variables is assumed. The polarization operator
$A^{ij}_J(r,r'|\omega)$ can be calculated in terms of the Green
function $G_{l j}(r,r'|\varepsilon)$ of the radial Schr\"{o}dinger
equation:
\begin{eqnarray}
\nonumber A^{ij}_J(r,r'|\omega)\,=\,\frac{1}{2J+1}\sum_{j_1 j_2 l_1 l_2}\,
t^{(i)}_{\nu_1\nu_2} t^{(j)}_{\nu_1\nu_2} \,
\sum_{n1} k_{\nu_1} R_{\nu_1}(r) R_{\nu_2}(r')\\
\times \left[ G_{l_2 j_2}(r,r'|\varepsilon_{\nu_1}-\omega)+G_{l_2
j_2}(r,r'|\varepsilon_{\nu_1}+\omega) \right],
\end{eqnarray}
where $k_{\nu}$ is the occupation number of the level $\nu=(n l j
m)$, and $R_{\nu}(r)$ is the single-particle radial wave function.
The Green function can be calculated using two linear independent
solution of the radial Schr\"{o}dinger equation. This method
allows one to use all of the single-particle spectrum
\cite{shlom},\cite{sap}. For the spin-orbit interaction
(\ref{lsint}) expression $\Theta^{ij}_{ls}$ in terms of $A^{ij}_J$
is more complicated (see \cite{dmit}).

\section{Hyperfine Splitting and Static MJ-moments}

The ground-state hyperfine splitting of hydrogen-like ions is
conveniently written in the form \cite{shab2}, \cite{tom1}
\begin{equation}\label{dE1}
\Delta E_{hfs}\,=\,\frac{4}{3}\alpha(\alpha Z)^3
\frac{\mu}{\mu_N}\frac{m_e}{m_p} \frac{2I+1}{2I}\,m_e c^2 \,
 A(\alpha Z)(1-\delta)(1-\varepsilon) + \Delta E_{QED}.
\end{equation}
Here $\alpha$ is the fine-structure constant, $Z$ is the nuclear
charge, $m_e$ is the electron mass, $m_p$ is the proton mass,
$\mu$ is the nuclear magnetic moment, $\mu_N$ is the nuclear
magneton, and $I$ is the nuclear spin. $A(\alpha Z)$ denotes the
relativistic factor \cite{breit}
\begin{equation}
A(\alpha Z)\,=\,\frac{1}{\gamma(2\gamma-1)},\; \mbox{where} \;
\gamma\,=\,\sqrt{1-(\alpha Z)^2}.
\end{equation}
$\delta$ is the nuclear charge distribution correction,
$\varepsilon$ is the nuclear magnetization distribution correction
(the Bohr-Weisskopf correction) \cite{bweis}, and $\Delta E_{QED}$
is the QED correction. To obtain the first part of Eq.
(\ref{dE1}), one can start from the magnetic interaction of an
electron and a nucleus
\begin{equation}\label{magnint}
H_{int}\,=\,-\frac{1}{c^2} \int d^3r d^3r'\frac{{\bf j}({\bf r})
\cdot {\bf J}({\bf r'})} {|\bf r - \bf r'|},
\end{equation}
where ${\bf j}({\bf r})$ and ${\bf J}({\bf r'})$ are the
electromagnetic current densities. The hfs can be written as
\begin{equation}\label{dE2}
\Delta E_{hfs}\,=\,\Delta E_1 + \Delta E_{QED},
\end{equation}
where $\Delta E_1$ is the diagonal matrix element of 0-th spherical
component of the vector vertex $\Gamma_{1 \mu}({\bf r})$,
\begin{equation}
\Delta E_1\,=\,\left< \nu m=I|\Gamma_{1 0}|\nu m=I\right>.
\end{equation}
The vertex $\Gamma_{1 \mu}$ is the solution of equation
(\ref{main}) with the bare vertex $\Gamma^0_{1 \mu}$ derived from
(\ref{magnint}),
\begin{eqnarray} \label{gamma0}\nonumber
\Gamma^0_{1 \mu}\,=\,\frac{4eK}{j+1}\,\frac{2I+1}{2I}\, \left\{
\mathcal{M}^0_{1 \mu} \Omega^{(0)}_{nlj}-\sqrt{4\pi}\,\mu_N
\sum_{a}\,\left[g_s(a) \, T^{(1)}_{1 \mu}(a)+g_l(a) \, T^{(3)}_{1
\mu}(a)
\right]\,\Omega^{(1)}_{nlJ}(r_a)\,\right.\\
\left.+\sqrt{4\pi}\,\mu_N \sum_{a}\,\left[g_l(a) \, T^{(3)}_{1
\mu}(a)\right.-\sqrt{\frac{1}{2}}\left. g_s(a) \, T^{(2)}_{1
\mu}(a) \right]\, \Omega^{(2)}_{nlJ}(r_a) \right\}.
\end{eqnarray}
In this expression the sum is taken over the nucleons, $K= (l-j)
(2j+1)$ for the atomic states with $j=l \pm \frac{1}{2}$,
\begin{eqnarray}
\Omega^{(0)}_{nlj}\,=\,\int_0^\infty d r' f_{nlj}(r')
g_{nlj}(r'),\\
\Omega^{(1)}_{nlj}(r)\,=\,\int_0^r d r' f_{nlj}(r')
g_{nlj}(r'),\\
\Omega^{(2)}_{nlj}(r)\,=\,\frac{1}{r^3}\,{\int}^r_0 d r' {r'}^3
f_{nlj}(r') g_{nlj}(r'),
\end{eqnarray}
where $f_{nlj}(r)$ and $g_{nlj}(r)$ are the upper and the lower
components of the Dirac radial wave function. The vertex
$\mathcal{M}^0_{1 \mu}$ is the bare operator of the nuclear
magnetic dipole moment. The bare operator of the nuclear magnetic
multipole moments has a standard form \cite{bm}
\begin{equation} \label{magmj}
\mathcal{M}^0_{JM}\,=\,\sqrt{4 \pi J} \mu_N \sum_a
\,r_a^{J-1}\,\left(g_s(a)\,T^{(1)}_{JM}(a) +
g_l(a)\,\frac{2}{J+1}T^{(3)}_{JM}(a)\right).
\end{equation}
The vertex $\Gamma^0_{1 \mu}$ is $M1$ part of the interaction
(\ref{magnint}) averaged over atomic state with the quantum
numbers $n$,$l$,$j$. Higher moments of the interaction
(\ref{magnint}) do not contribute to hfs for the electron ground
state with $j=1/2$.

The same equations (\ref{main}) can be used to calculate nuclear
magnetic multipole moments $\mathcal{M}_{JM}$ with the bare vertex
(\ref{magmj}). The Eqs. (\ref{gamma0}), (\ref{magmj}) obtained
using the standard electromagnetic current density for free
nucleons. However, due to velocity dependence of the spin-orbit
interaction (\ref{lsint}), there are the corrections to
electromagnetic current density. The corrections were discussed
earlier \cite{bm,dmit}. They produce an additional contributions
to the bare vertexes $\mathcal{M}^0_{JM}$ and $\Gamma^0_{1\mu}$
and can be presented in the similar form as
(\ref{gamma0}),(\ref{magmj}). The corresponding equations can be
found in the Appendix A.

It is worth to note, that the expression for hfs (\ref{gamma0}) is
a sum of three terms. The first, being proportional
$\Omega^{(0)}_{nlj}$, corresponds to a point-like magnetic moment
distribution. It does not contribute to the correction $\epsilon$.
This term gives the main contribution to hfs being larger than two
other terms by about two orders of magnitude. The effects of
nuclear structure are not essential here, they are hidden in the
value of the nuclear magnetic moment which is well known from the
experiment. Two other terms are, just, the correction for an
extended nuclear magnetization. But, only the second term is
proportional to the magnetic moment density, while the third one
is not.

The kernel (\ref{kern}) of the Eq.(\ref{main2}) was calculated
using partially self-consistent mean field potential \cite{birb}.
The potential includes four terms. The isoscalar term is  the
standard Woods-Saxon potential
\begin{equation} \label{ws}
U_0(r)=-\frac{V}{1+\exp{\frac{r-R}{a}}},
\end{equation}
with the parameters  $V=52.03$ MeV, $R=1.2709A^{1/3}$ fm, and
$a=0.742$ fm \cite{birb}. Two other terms $U_{ls}(r)$, and
$U_\tau(r)$ were calculated
 self-consistently using two-body interaction Eq.(\ref{lsint}) for the spin-
orbit part of the potential, and the Migdal-type isovector interaction
\begin{equation} \label{tayint}
F_\tau = f'C(\mbox{\boldmath $\tau$}_1\cdot\mbox{\boldmath $\tau$}_2)
\delta ({\bf r}_1-{\bf r}_2)
\end{equation}
for the isovector part of the potential, with $f'=1.075$
\cite{birb}. The last term is the Coulomb interaction that was
calculated for uniform charge sphere distribution with $R_C=1.18
A^{1/3}$. This potential produces for $^{209}$Bi nucleus a charge
density with rms. $\left< r_c{}^2 \right>^{1/2} = 5.52$ fm, which
is in fair agreement with the measured value 5.519(4) fm
\cite{devri}.

The parameters of the interactions Eqs.(\ref{ssint},\ref{lsint}) used in
calculations are listed in Table 1. This is a standard set of the parameters
used in lead region \cite{sap}.
\renewcommand{\arraystretch}{1.5}
\begin{table}
\caption{Interaction parameters}
\begin{center}
\begin{tabular}{|c|c|c|c|c|c|}
\hhline{|------|}
$C$ (MeV$\cdot$ fm$^3$)&$r_0$ (fm)&$\kappa$&$\kappa'$&$g$&$g'$ \\
\hhline{|------|}
300 & 1.16 & 0.21 & -0.12 & 0.633 & 1.0133 \\
\hhline{|------|}
\end{tabular}
\end{center}
\end{table}
\renewcommand{\arraystretch}{1.}
\section{Effects of the core polarization}
Qualitatively, the effects of the core polarization are well
understood. The spin-spin interaction is repulsive, therefore, it
creates negative core response decreasing the spin contribution to
the magnetic moment. Fig. 1 shows the core polarization effects on
the spin part of the magnetic moment in ${}^{209}$Bi. The minimum
near nuclear surface is due to transitions to the partially filled
upper level $1h_{9/2}$ of the spin-orbit doublet. The transitions
to higher states, including continuum, produce a uniform decrease
of $g_s$ near $r=0$.
\begin{figure}
\includegraphics[width=0.47\textwidth]{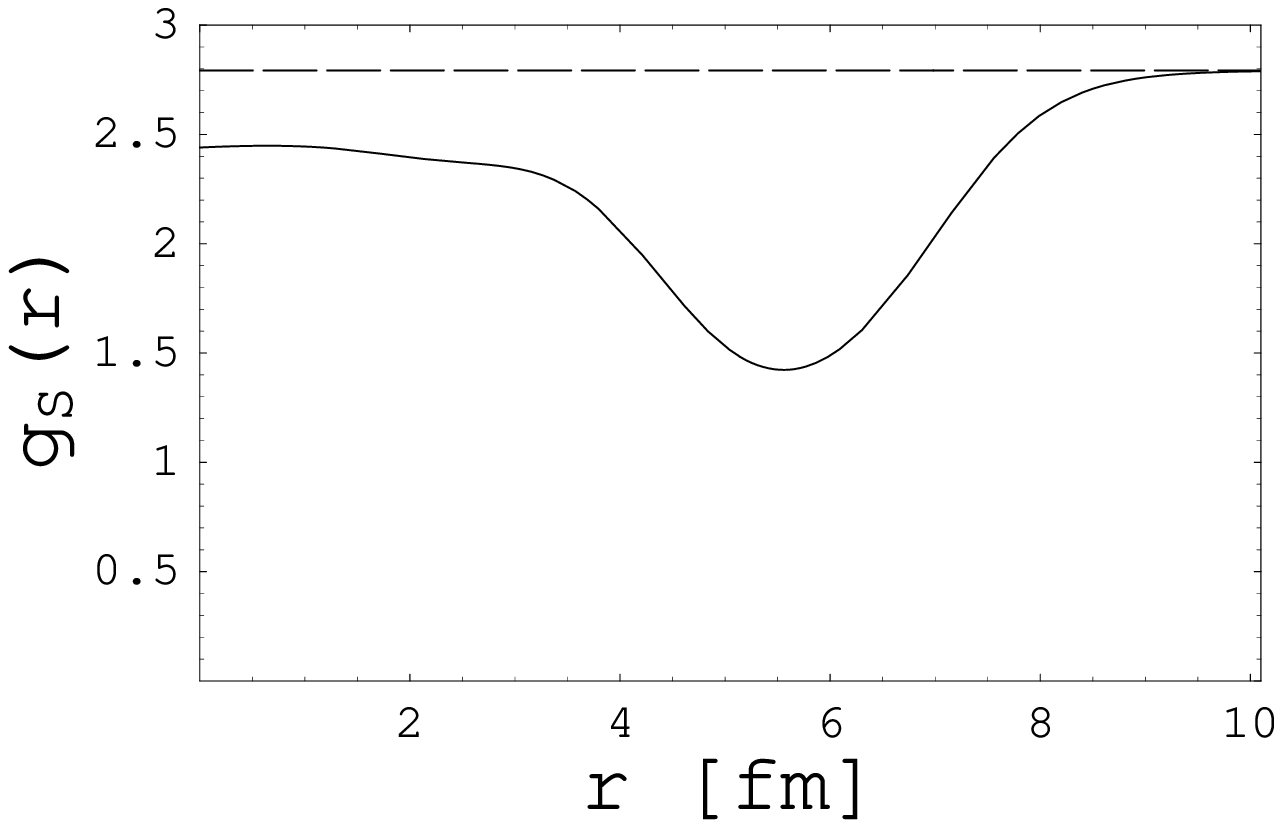}
\hfill
\includegraphics[width=0.48\textwidth]{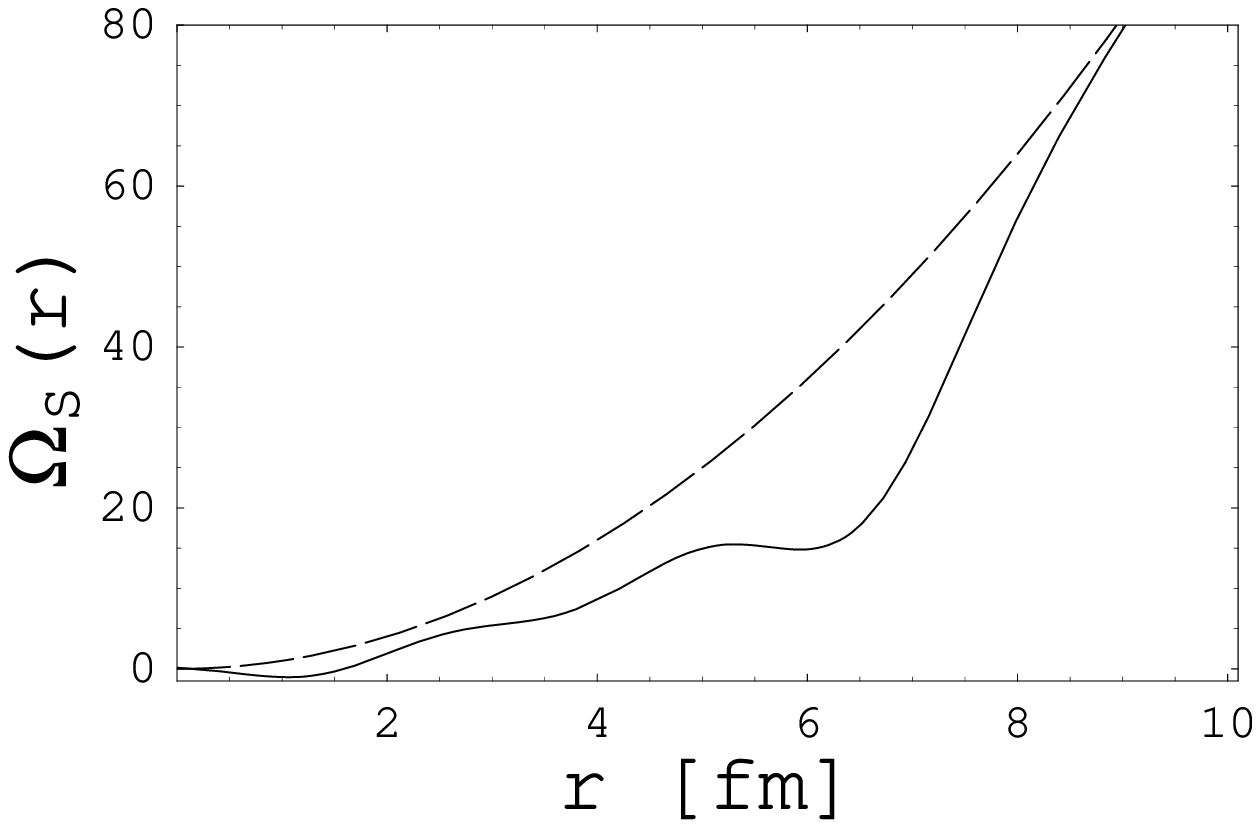}
\\
\parbox[t]{0.47\textwidth}{\caption{Effects of the core polarization on the spin part of the
magnetic moment in the ${}^{209}$Bi. Dashed line shows
unrenormalized $g_s=2.79$. Full line shows renormalized $g_s$ as a
function of $r$.}} \hfill
\parbox[t]{0.48\textwidth}{\caption{Effects of the core polarization on the spin part of the
octupole moment in ${}^{209}$Bi. Dashed line shows unrenormalized
$\Omega_s=r^2$ dependence. Full line shows renormalized $\Omega_s$
as a function of $r$.}}
\end{figure}
The core polarization effects are more pronounced in the octupole
moment of ${}^{209}$Bi. Fig. 2 show how the original $r^2$
dependence is modified by the transitions over Fermi surface. The
number of the transitions is greater than in case of magnetic
dipole moment. Here the transitions with $\Delta N=2$ are involved
as well, therefore, the relative effect of the polarization is
greater than for the magnetic moment.

Although the core polarization increases absolute values both of
the magnetic moment and the octupole moment, this increase is not
enough to explain the data. Table 2. shows the contributions to
the magnetic dipole and octupole moments in ${}^{209}$Bi from
$\mathcal{M}^0$ and $\Delta\mathcal{M}^0$ and the effects of the
core polarization when $\zeta_s = \zeta_l =0$.
\begin{table}
\begin{center}
\begin{tabular}{|c|c|c|c|c|c|c|}
\hline
&$\mathcal{M}^0$&$\Delta\mathcal{M}^0$&$\mathcal{M}$&$\Delta\mathcal{M}$&$
\mathcal{M}+\Delta\mathcal{M}$& Exp.\\
\hline
dipole&2.62&0.49&3.28&0.30&3.58&4.110(4)\footnotemark[1] \\
\hline
octupole&-14&-9&-44&-4&-48&-55(3)\footnotemark[2] \\
\hline
\end{tabular}
\parbox[t]{0.60\textwidth}{\caption{Contributions to the magnetic dipole and octupole
moments in the ${}^{209}$Bi,  $\mu_N \cdot $fm${}^{J-1}$.}}
\end{center}
\end{table}
\footnotetext[1]{The value was taken from Ref.\cite{mm}}
\footnotetext[2]{The value was taken from Ref.\cite{Lan}} The
remaining contribution comes from higher orders configuration
mixing, and from other possible velocity dependent forces not
included into consideration. This contribution is expected to be
small. It can be taken into account phenomenologically, via the
effective charges $\zeta_s$ and $\zeta_l$. Fitting the value of
the ${}^{209}$Bi magnetic moment only is not enough to find both
$\zeta_s$ and $\zeta_l$. To obtain the region of the allowed values
for $\zeta_s$ and $\zeta_l$ we calculated the magnetic moments for
all four nuclei lying near doubly-magic nucleus ${}^{208}$Pb. They
are ${}^{207}$Tl,${}^{207}$Pb,${}^{209}$Bi, and ${}^{209}$Pb. Our
suggestion is that for these four nuclei the difference between
their effective charges is small and can be neglected. A
calculated magnetic moment is a linear function of the
giromagnetic ratios $g_s,g_l$. So, on the $\zeta_s,\zeta_l$ plane,
all points where the calculated magnetic moment is equal to its
measured value, lie on a straight line. In Fig. 3 we
plotted the lines corresponding to fits of the magnetic moments for
the set of nuclei, mentioned above. The data were taken from Ref.\cite{stone}.
The accuracy of the
measurement is high, and two lines corresponding to the
experimental values $\mu \pm \Delta \mu$ almost coincide within
the scale of the figure. In an ideal theory all four lines should
cross in the same point. This is not the case for our model. The allowed values
for $\zeta_s$ and $\zeta_l$ form the whole region which is shown in Fig.3 by
the shaded triangle. The ``best'' values of $\zeta_s$ and $\zeta_l$ were found
by minimization of a mean squared deviation of the calculated magnetic
dipole moments for these four nuclei from the data. They are $\zeta_l=-0.074,
\zeta_s=0.030$. These values are small enough in line with the above
suggestion.
The
size of the shaded triangle in Fig. 3 can be used to estimate the
accuracy of our approach. In our model the uncertainties in calculation of
the hfs were found by varying
$\zeta_s$ and $\zeta_l$ within this shadow triangle.
\begin{figure}
\centering
\includegraphics[width=0.5\textwidth]{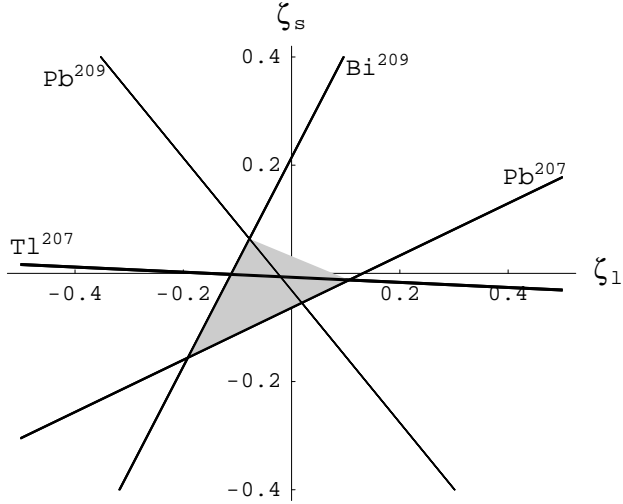}
\\
\parbox[t]{\textwidth}{\caption{Region of allowed values for $\zeta_l$ and
$\zeta_s$ with core polarization effects.}}
\end{figure}

The resulting distribution of the magnetic moment density in
${}^{209}$Bi is shown in Fig. 4. The magnetization density is
peaked near nuclear surface both for the bare contribution of the
unpaired proton, shown by a dashed line, and for the total density
shown by full line. The peak position in the total magnetization density
is shifted slightly to smaller {\it r}. This is related directly to the
decrease of $g_s$ inside the nucleus (see Fig. 1). The decrease leads to
enhancement of the magnetic moment in this region due to opposite sign
of the spin part of the magnetic moment relative to its orbital contribution
for $h_{9/2}$ level. The shift of the peak results in some decrease of the
magnetization rms-radius.
The calculated magnetization rms-radius is
practically insensitive to the values of $\zeta_l$ and $\zeta_s$
within the above range. It is equal to $\langle r_m^2
\rangle^{1/2} = 5.86$ fm.
Fig. 5 gives similar figure for the
octupole moment density distribution in ${}^{209}$Bi. In Figs. 4,5
one can see that the core polarization effects for the
octupole moment are really larger than for the magnetic moment.
\begin{figure}
\includegraphics[width=0.47\textwidth]{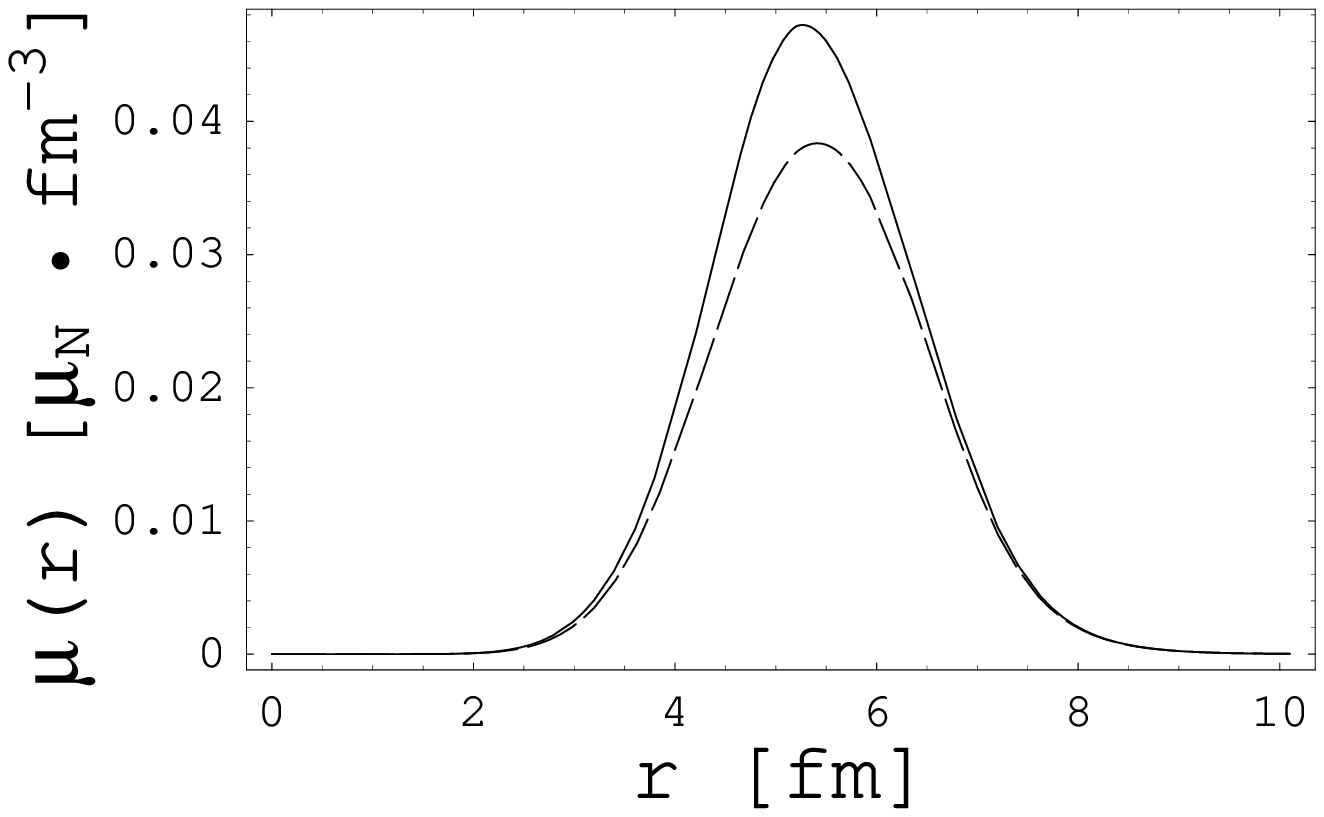}
\hfill
\includegraphics[width=0.47\textwidth]{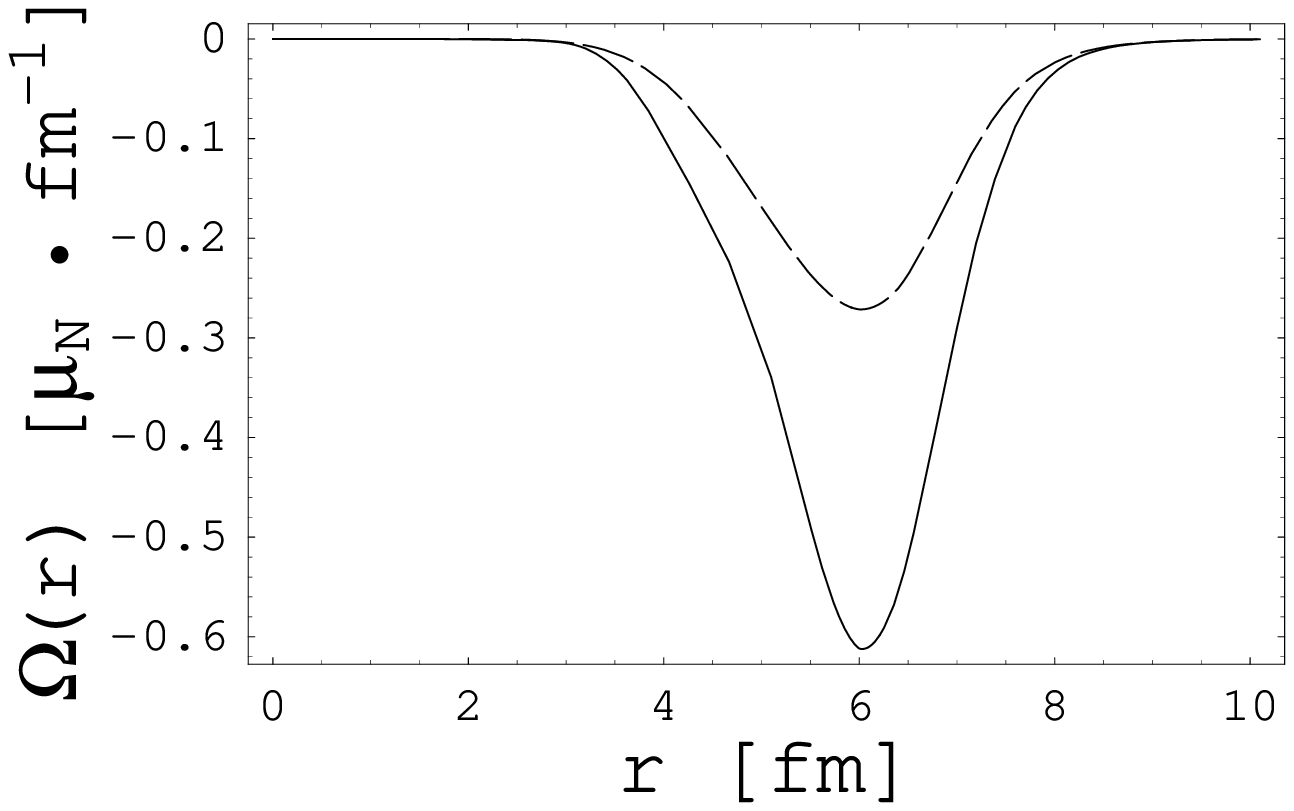}
\\
\parbox[t]{0.47\textwidth}{\caption{Magnetic moment density distribution in
${}^{209}$Bi. A contribution from a bare unpaired proton is shown
by a dashed line. Full line shows the total magnetic moment
density including core polarization.}} \hfill
\parbox[t]{0.47\textwidth}{\caption{Octupole moment density distribution in
${}^{209}$Bi. A contribution from a bare unpaired proton is shown
by a dashed line. Full line shows the total octupole moment
density including core polarization.}}
\end{figure}
\begin{table}
\caption{Theoretical values of the ground-state hfs and the
experimental results. The major contribution to uncertainties of
the theoretical values originates from the uncertainties in the
Bohr-Weisskopf effect.}
\begin{center}
\begin{tabular}{|l|l|l|l|l|l|l|}
\cline{1-3}\cline{5-7} \multicolumn{3}{|c|}{\slshape } &
\hspace{1mm} & \multicolumn{3}{|c|}{\slshape }\\
$E_{hfs}$ in eV &${}^{209}$Bi${}^{82+}$&${}^{207}$Pb${}^{81+}$& & &$\epsilon$ for ${}^{209}$Bi${}^{82+}$&$\epsilon$ for ${}^{207}$Pb${}^{81+}$\\
\cline{1-3}\cline{5-7} $E_0$&5.191(5)&1.274(3)& &Our work &
0.0095$(\!\!\!{\scriptsize
\begin{array}{c}
+7 \\ -38
\end{array}}\!\!\!)$& 0.0353$(\!\!\!{\scriptsize
\begin{array}{c}
-35 \\ +164
\end{array}}\!\!\!)$\\
$\Delta E_{BW}$& -0.050$(\!\!\!{\scriptsize
\begin{array}{c}
-3 \\ +20
\end{array}}\!\!\!)$&-0.045$(\!\!\!{\scriptsize
\begin{array}{c}
+4 \\ -21
\end{array}}\!\!\!
)$& &Shabaev et al. \cite{shab2} & 0.0118& 0.0419\\
$\Delta E_{QED}$&-0.030&-0.007& &Gustavsson et al. \cite{gust} & 0.0131(26)& 0.0429(86)\\
$E_{hfs}^{theory}$&5.111$(\!\!\!{\scriptsize
\begin{array}{c}
-3 \\ +20
\end{array}}\!\!\!
)$(5)&1.222$(\!\!\!{\scriptsize
\begin{array}{c}
+4 \\ -21
\end{array}}\!\!\!
)$(3)& &Tomaselli et al. \cite{tom2} & 0.0210(17)& 0.0289(15)\\
$E_{hfs}^{exper.}$&5.0840(8)\footnotemark[3]&1.2159(2)\footnotemark[4]& &Experiment&0.0147(11)&0.0397(25)\\
\cline{1-3}\cline{5-7}
\end{tabular}
\end{center}
\end{table}
\footnotetext[3]{The value was taken from Ref.\cite{hfs}}
\footnotetext[4]{The value was taken from Ref.\cite{hfs2}}

The results
obtained for hfs are listed in the left part of Table 3. Here
$E_0$ corresponds to the relativistic Fermi formulae plus charge
distribution correction. The uncertainty in the value of $E_0$
originates mainly from the uncertainty of the measured magnetic
moment: $\mu=4.110(4) \mu_N$ for ${}^{209}$Bi and $\mu=0.5925(6)
\mu_N$ for ${}^{207}$Pb \cite{mm}. The uncertainty of the charge
distribution correction is small and can be neglected. The
uncertainties in Bohr-Weisskopf effect $\Delta E_{BW}$ come from
the uncertainties in the effective charges $\zeta_s$ and $\zeta_l$.
They were determined using the shadow triangle shown in Fig. 3. They
are not symmetric since the ``best'' values of $\zeta_l$ and $\zeta_s$
are lying close to $^{209}$Bi line in Fig. 3. In
order to compare our calculations with the data we need $\Delta
E_{QED}$. It was calculated in several papers and the results
are consistent \cite{shab2,schn,perss,blund,pers}. In \cite{pers}
it was shown that $\Delta E_{QED}$ is insensitive to details of
the magnetization distribution. They give for $\Delta E_{QED}$ the
value $\Delta E_{QED}= -0.0298(3)$ eV for ${}^{209}$Bi${}^{82+}$
and $\Delta E_{QED}= -0.00726(7)$ eV for ${}^{207}$Pb${}^{81+}$.
In our calculated values of the ground-state hfs
$E_{hfs}^{theory}$ the uncertainty in the first brackets
originates from $\Delta E_{BW}$ and the second one originates from
$\Delta E_{QED}$.
The right part of the Table 3 gives our results for Bohr-Weisskopf
effect in comparison with some other theoretical calculations.
While our result for the ground-state hfs of
$^{207}$Pb$^{81+}$ is in good agreement with the data, the result
for $^{209}$Bi$^{82+}$ lies out of the data. The calculated $\epsilon$
at its upper limit $\epsilon=0.0103$ differs by  four standard
deviations from the measured value $\epsilon = 0.0147 \pm 0.0011$. Actually,
$\epsilon$ is not directly measured value. It was obtained from the measured
$E^{exper.}_{hfs}$ using the calculated values of $E_0$ and $\Delta E_{QED}$.
The uncertainty in $\epsilon$ comes mainly from the uncertainties in these
calculated $E_0$.
In other calculations only in \cite{tom2} the particle-phonon coupling
was accounted in scope of the DCM. In \cite{shab2}, and \cite{gust}, and
others (see Refs. in \cite{gust}) the unpaired particle was treated as
an independent particle moving in a mean field potential. In this approach,
the difference in $\epsilon$ reflects more the sensitivity to the choice
of a particular mean field potential. It is interesting to note that
$\epsilon$ calculated in \cite{tom2} lies almost on the same distance
from the experimental value, but on the other side compared to our value.
This difference may be attributed to ground state correlations that
contribute significantly to transition probabilities at small excitation
energies \cite{brown}. In order to check whether our results is sensitive
to a particular parameterization of the effective charges we made calculation
of the hfs in $^{209}$Bi$^{82+}$ treating all the giromagnetic ratios
$g_l(p), g_l(n), g_s(p),$ and $g_s(n)$ as free parameters. Their values were
obtained by fitting the magnetic moments of four nuclei mentioned above.
The magnetic moments were calculated together with the core polarization
effects and the $ls$-corrections. With these new parameters we calculated
hfs for $^{209}$Bi$^{82+}$ and obtained exactly the same result $\Delta
E_{BW} = -0.050$ eV. This situation seems to be rather general. We changed
different parameters entering in our theory, including $g$ and $g'$, although
$g'$ is fixed by the position of Gamow-Teller resonances \cite{ost}. But, as
soon as the magnetic moment of $^{209}$Bi is fitted, the correction $\Delta
E_{BW}$ becomes close to the value cited in Table 3.

In addition to the magnetic moment, for $^{209}$Bi there are other observables
related to the magnetic properties. In Table 4 we summarized our results
including there our calculations of the octupole moment of $^{209}$Bi and
the hfs in muonic $^{209}$Bi. To demonstrate the relative importance of the
core polarization effects for different observables we made additional
calculation switching off the core polarization and keeping the same
parameterization of the effective charges via $\zeta_l$ and $\zeta_s$.
One can see that although the effects of the core polarization are not very
significant for the electronic hfs, for other observables, like the octupole
moment, they are rather large. All the observables except
the electronic hfs in $^{209}$Bi are in good agreement with the data. However,
the experimental uncertainties in case of the octupole moment and the muonic
hfs are rather large. It would be very desirable to reduce them in order to
see whether the discrepancy in the electronic hfs in $^{209}$Bi were
pronounced in these observables as well at higher accuracy of the data.

In summary, using continuum RPA with effective residual forces we
calculated the distribution of the magnetic and the octupole densities for
$^{209}$Bi.
Additional contribution from velocity dependent spin-orbit
two-particle interaction was calculated explicitly. The parameters of the
theory were fixed by fitting the magnetic moments of four one-particle and
one-hole nuclei around $^{208}$Pb.
Basing on
these results we calculated the hfs in hydrogen-like
$^{209}$Bi$^{82+}$ and $^{207}$Pb$^{81+}$ ions, the octupole
magnetic moment in $^{209}$Bi, and hfs in muonic atom of
$^{209}$Bi. Except the electronic hfs in hydrogen-like $^{209}$Bi$^{82+}$
all other
results are in good agreement with experiment.

\section{Acknowledgments}
The authors appreciate the discussions with I.B. Khriplovich.
\begin{table}
\caption{Relative importance of the  core polarization effects for
different observables.}
\begin{center}
\begin{tabular}{|l|l|l|l|}
\hline
&Without core p. effects&With core p. effects&Experiment\\
\hline $\Delta E_{BW}$ for ${}^{209}$Bi,
eV&-0.046$(\!\!\!{\scriptsize
\begin{array}{c}
-3 \\ +39
\end{array}}\!\!\!)$&
-0.050$(\!\!\!{\scriptsize
\begin{array}{c}
-3 \\ +20
\end{array}}\!\!\!)$&\\
$E_{hfs}$ for ${}^{209}$Bi, eV&5.115$(\!\!\!{\scriptsize
\begin{array}{c}
-3 \\ +39
\end{array}}\!\!\!)$(5)&5.111$(\!\!\!{\scriptsize
\begin{array}{c}
-3 \\ +20
\end{array}}\!\!\!)$(5)&5.0840(8)\\
$\Delta E_{BW}$ for ${}^{207}$Pb, eV&-0.035$(\!\!\!{\scriptsize
\begin{array}{c}
+3 \\ -37
\end{array}}\!\!\!
)$&-0.045$(\!\!\!{\scriptsize
\begin{array}{c}
+4 \\ -21
\end{array}}\!\!\!
)$&\\
$E_{hfs}$ for ${}^{207}$Pb, eV&1.232$(\!\!\!{\scriptsize
\begin{array}{c}
+3 \\ -37
\end{array}}\!\!\!
)$(3)&1.222$(\!\!\!{\scriptsize
\begin{array}{c}
+4 \\ -21
\end{array}}\!\!\!
)$(3)&1.2159(2)\\
$\Delta E_{BW}$ for muonic ${}^{209}$Bi,
KeV&-1.73$(\!\!\!{\scriptsize
\begin{array}{c}
-9 \\ +129
\end{array}}\!\!\!
)$&-1.86$(\!\!\!{\scriptsize
\begin{array}{c}
-10 \\ +64
\end{array}}\!\!\!
)$&\\
$E_{hfs}$ for muonic ${}^{209}$Bi, KeV&4.76$(\!\!\!{\scriptsize
\begin{array}{c}
-9 \\ +129
\end{array}}\!\!\!
)$(6)&4.63$(\!\!\!{\scriptsize
\begin{array}{c}
-10 \\ +64
\end{array}}\!\!\!
)$(6)&4.44(15) \footnotemark[5]\\
Octupole moment of ${}^{209}$Bi, $\mu_N
\cdot$fm${}^2$&-42$(\!\!\!{\scriptsize
\begin{array}{c}
-2 \\ +30
\end{array}}\!\!\!
)$&-55$(\!\!\!{\scriptsize
\begin{array}{c}
-2 \\ +16
\end{array}}\!\!\!
)$&-55(3)\\
\hline
\end{tabular}
\end{center}
\end{table}
\footnotetext[5]{The value was taken from Ref. \cite{R86}}
\appendix
\section{Spin-orbit corrections to the bare vertices}
The vertex $\mathcal{M}_{JM}$ is the solution of equation
(\ref{main}) with the bare vertex (\ref{magmj}). The Eqs.
(\ref{gamma0}), (\ref{magmj}) obtained using the standard
electromagnetic current density for free nucleons. However, due to
velocity dependence of the spin-orbit interaction (\ref{lsint}),
there are the corrections to electromagnetic current density. The
corrections were discussed  earlier \cite{bm,dmit}. They produce
an additional contributions to the bare vertexes
$\mathcal{M}^0_{JM}$ and $\Gamma^0_{1\mu}$,
$$
\Delta \mathcal{M}^{0p}_{JM} =\sqrt{4\pi J} \mu_N
\frac{2mr^2_0}{\hbar^2}C\sum_a \left[-(\kappa + \kappa')
r^{J-1}_a\rho_p(r_a)+(\kappa - \kappa')\frac{1}{2J+1}r^J_a
\frac{d\rho_n(r_a)}{dr}\right]T^{(1)}_{JM}(a)
$$
\begin{equation} \label{corrmjp}
 +(\kappa - \kappa')\frac{1}{(2J+1)}\sqrt{\frac{J}{J+1}}r^J_a
\frac{d\rho_n(r_a)}{dr}T^{(2)}_{JM}(a),
\end{equation}
$$
\Delta \mathcal{M}^{0n}_{JM} =\sqrt{4\pi J} \mu_N
\frac{2mr^2_0}{\hbar^2}C\sum_a \left[-(\kappa - \kappa')
r^{J-1}_a\rho_p(r_a)-(\kappa - \kappa')\frac{1}{2J+1}r^J_a
\frac{d\rho_p(r_a)}{dr}\right]T^{(1)}_{JM}(a)
$$
\begin{equation} \label{corrmjn}
 -(\kappa - \kappa')\frac{1}{(2J+1)}\sqrt{\frac{J}{J+1}}r^J_a
\frac{d\rho_p(r_a)}{dr}T^{(2)}_{JM}(a).
\end{equation}
$$
\Delta \Gamma^{0p}_{1\mu}=\frac{4eK}{j+1}\frac{2I+1}{2I}\left\{
\Delta \mathcal{M}^{0p}_{1 \mu}\,\Omega^{(0)}_{nlj}-\sqrt{4\pi}
\mu_N \frac{2 m r^2_0}{\hbar^2} C \sum_a\left[-\left( (\kappa
+\kappa')\rho_p(r_a)- \frac{1}{3}(\kappa -
\kappa')r_a\frac{d\rho_n(r_a)}{dr}\right)
\Omega^{(1)}_{nlj}(r_a)\right.\right.
$$
$$
\left.-\frac{1}{3}(\kappa - \kappa')r_a\frac{d\rho_n(r_a)}{dr}
\Omega^{(2)}_{nlj}(r_a) \right] T^{(1)}_{1\mu}(a) +\left[
\frac{1}{3}(\kappa -\kappa')r_a \frac{d\rho_n(r_a)} {dr}
\Omega^{(1)}_{nlj}(r_a)\right.
$$
\begin{equation} \label{corrgp}
\left.\left.-\left( (\kappa +
\kappa')\rho_p(r_a)+\frac{1}{3}(\kappa -
\kappa')r_a\frac{d\rho_n(r_a)}{dr}\right)
\Omega^{(2)}_{nlj}(r_a)\right]\sqrt{\frac{1}{2}}T^{(2)}_{1\mu}(a)\right\},
\end{equation}
$$
\Delta \Gamma^{0n}_{1\mu}=\frac{4eK}{j+1}\frac{2I+1}{2I}\left\{
\Delta \mathcal{M}^{0n}_{1 \mu}\,\Omega^{(0)}_{nlj}-\sqrt{4\pi}
\mu_N\frac{2 m r^2_0}{\hbar^2} C \sum_a\left[-\left( (\kappa
-\kappa') \rho_p(r_a) + \frac{1}{3}(\kappa -
\kappa')r_a\frac{d\rho_p(r_a)}{dr}\right)
\Omega^{(1)}_{nlj}(r_a)\right.\right.
$$
$$
\left.+\frac{1}{3}(\kappa - \kappa')r_a\frac{d\rho_p(r_a)}{dr}
\Omega^{(2)}_{nlj}(r_a) \right] T^{(1)}_{1\mu}(a) +\left[
-\frac{1}{3}(\kappa -\kappa')r_a \frac{d\rho_p(r_a)}{dr}
\Omega^{(1)}_{nlj}(r_a)\right.
$$
\begin{equation} \label{corrgn}
\left.\left.+\left( -(\kappa -
\kappa')\rho_p(r_a)+\frac{1}{3}(\kappa -
\kappa')r_a\frac{d\rho_p(r_a)}{dr}\right)
\Omega^{(2)}_{nlj}(r_a)\right]\sqrt{\frac{1}{2}}T^{(2)}_{1\mu}(a)\right\}.
\end{equation}

\end{document}